\documentclass[preprint,10pt]{article}
\usepackage{cite}
\usepackage{amsmath,amssymb,amsfonts}
\usepackage{algorithmic}
\usepackage{graphicx}
\usepackage{textcomp}
\usepackage{listings}
\usepackage{color}
\usepackage{authblk}

\newcommand\mnewcommand[1]{%
\let#1\relax \newcommand#1 }

\mnewcommand{\diver}{\mnabla \cdot}
\mnewcommand{\vel}{\mathbf{u}}
\mnewcommand{\velhg}{\vel_h}
\mnewcommand{\preshg}{{p}_h}
\mnewcommand{\dim}{3}
\mnewcommand{\ud}{d}
\mnewcommand{\vort}{\mathbf{w}}
\mnewcommand{\rot}[1]{\mnabla \times #1}
\mnewcommand{\selfinnerprod}[1]{\innerprod{#1}{#1}}
\mnewcommand{\innerprod}[2]{< #1 , #2 >}
\mnewcommand{\lapl}{\Delta}
\mnewcommand{\Nx}{N_1}
\mnewcommand{\Ny}{N_2}
\mnewcommand{\Nz}{N_3}
\mnewcommand{\Nm}{N_m}
\mnewcommand{\Ns}{N_s}
\mnewcommand{\complconj}[1]{#1^{*}}
\mnewcommand{\step}{\Delta}
\mnewcommand{\dt}{\step t}
\mnewcommand{\traspose}{^{*}}
\mnewcommand{\avg}[1]{\overline{#1}}
\mnewcommand{\mcdot}{\bcdot}
\mnewcommand{\mnabla}{\nabla}
\mnewcommand{\real}{\mathbb{R}}
\mnewcommand{\complex}{\mathbb{C}}

\mnewcommand{\matvec}[1]{\mathbf{#1}}
 \mnewcommand{\mA}{\matvec{A}}
\mnewcommand{\mx}{\matvec{x}}
\mnewcommand{\mb}{\matvec{b}}

\providecommand\bnabla{\boldsymbol{\nabla}}

\mnewcommand{\mnabla}{\bnabla}
\mnewcommand{\lapl}{\mnabla^2}
\mnewcommand{\diver}{\mnabla \cdot}
\mnewcommand{\vel}{\mathbf{u}}
\mnewcommand{\vels}{u_s}
\mnewcommand{\gradvel}{\mathbf{g_u}}
\mnewcommand{\gradp}{\mathbf{g}_c}
\mnewcommand{\presh}{p_h}
\mnewcommand{\presc}{p_c}
\mnewcommand{\pres}{p}
\mnewcommand{\nuh}{\mathbf{\nu}_h}
\mnewcommand{\nut}{\nu_{t}}
\mnewcommand{\massf}{m_f}
\mnewcommand{\massfik}{\mathbf{m}_f_{ik}}
\mnewcommand{\bodyforce}{\boldsymbol{f}}

\def\BibTeX{{\rm B\kern-.05em{\sc i\kern-.025em b}\kern-.08em
    T\kern-.1667em\lower.7ex\hbox{E}\kern-.125emX}}
\begin{document}

\title{An FPGA cached sparse matrix vector product (SpMV) for unstructured computational fluid dynamics simulations}
%\\

%\author{\IEEEauthorblockN{Guillermo Oyarzun\footnotesize \textsuperscript{*}} \thanks{\textsuperscript{*} Corresponding author.}
%\IEEEauthorblockA{\textit{Computer Applications in Science and Engineering } \\
%\textit{Barcelona Supercomputing Center}\\
%Barcelona, Spain \\
%guillermo.oyarzun@bsc.es}
%\and
%\IEEEauthorblockN{Daniel Peyrolon}
%\IEEEauthorblockA{\textit{Computer Sciences} \\
%\textit{Barcelona Supercomputing Center}\\
%Barcelona, Spain \\
%daniel.peyrolon@bsc.es}
%\and
%\IEEEauthorblockN{Carlos Alvarez}
%\IEEEauthorblockA{\textit{Computer Sciences} \\
%\textit{Barcelona Supercomputing Center}\\
%Barcelona, Spain \\
%carlos.alvarez@bsc.es}
%\and
%\IEEEauthorblockN{Xavier Martorell}
%\IEEEauthorblockA{\textit{Computer Sciences} \\
%\textit{Barcelona Supercomputing Center}\\
%Barcelona, Spain \\
%xavier.martorell@bsc.es}
%}

%\author{\IEEEauthorblockN{Guillermo Oyarzun \IEEEauthorrefmark{4} \textsuperscript{*}  \thanks{\textsuperscript{*} Corresponding author. guillermo.oyarzun@bsc.es},
%Daniel Peyrolon\IEEEauthorrefmark{2},
%Carlos Alvarez\IEEEauthorrefmark{2},
%Xavier Martorell\IEEEauthorrefmark{2}
%}
%\IEEEauthorblockA{\IEEEauthorrefmark{4}\textit{Computer Applications in Science and Engineering } \\
%\textit{Barcelona Supercomputing Center }\\
% Spain \\}
%\IEEEauthorblockA{\IEEEauthorrefmark{2}\textit{Computer Sciences} \\
%\textit{Barcelona Supercomputing Center }\\
% Spain \\
%}
%}

\author[1]{Guillermo Oyarzun \thanks{Corresponding author. guillermo.oyarzun@bsc.es}}
\author[2]{Daniel Peyrolon} 
\author[2]{Carlos Alvarez} 
\author[2]{Xavier Martorell} 

\affil[1]{Computer Applications in Science and Engineering, Barcelona Supercomputing Center }
\affil[2]{Computer Sciences, Barcelona Supercomputing Center }

\maketitle

\begin{abstract}
%This document is a model and instructions for \LaTeX.
%This and the IEEEtran.cls file define the components of your paper [title, text, heads, etc.]. *CRITICAL: Do Not Use Symbols, Special Characters, Footnotes, 
%or Math in Paper Title or Abstract.
%
Field Programmable Gate Arrays generate algorithmic specific architectures that improve the codes' FLOP per watt ratio. Such devices are re-gaining interest due to the rise of new tools that facilitate their programming, such as OmpSs. The computational fluid dynamics community is always investigating new architectures that can improve its algorithms' performance.  Commonly, those algorithms have a low arithmetic intensity and only reach a small percentage of the peak performance. The sparse matrix-vector multiplication is one of the most time-consuming operations on unstructured simulations. The matrix's sparsity pattern determines the indirect memory accesses of the multiplying vector.  This data path is hard to predict,  making traditional implementations fail.  In this work, we present an FPGA architecture that maximizes the vector's re-usability by introducing a cache-like architecture. The cache is implemented as a circular list that maintains the BRAM vector components while needed. Following this strategy, up to 16 times of acceleration is obtained compared to a naive implementation of the algorithm.
\end{abstract}

%\keywords{FPGA, SpMV, CFD, cache, energy efficient}

\section{Introduction}

Power constraints are pushing the High-Performance Computing (HPC) community towards the search for more efficient computing architectures that can cope with the exascale's challenges\cite{DON11}. One of the current trends consists of using accelerators as co-processors that increase the Floating-point operations (FLOP) per watt ratio in the systems. The Graphic Processor Units (GPU)s are the most common accelerators utilized in the leading edge supercomputers\cite{top500}. Its adoption has been supported by the impulse in markets such as videogames, cryptocurrency, and deep learning. Those markets are focused on dense computations that do not always reflect the algorithms' needs in computational fluid dynamics (CFD). Consequently, CFD codes are accelerated but can only exploit a small percentage of the peak performance\cite{OYA20, BOR20}. The results are even more dramatic when dealing with simulations based on unstructured meshes.  In that case, the parallel algorithms are characterized by sparse algebra operations and low arithmetic intensity.

Field Programmable Gate Arrays (FPGAs) are accelerators capable of reconfiguring their circuit logic. The chip consists of an array of configurable logic blocks interconnected by a programmable network. Those logic blocks are supported by an internal memory called r Block Random Access Memory (BRAM).  Also, the blocks can include non-configurable computing engines dedicated to specific tasks. The FPGAs' advantage is their ability to use problem-specific information to create customized architectures that efficiently map the data path and algorithmic logic. By doing so, the chip is entirely devoted to specific tasks, increasing the power efficiency in comparison with general-purpose architectures (CPUs and GPUs). For years, the developers viewed the FPGAs as complex devices that require in-depth computing architecture knowledge with unreasonable long developing cycles. This fact has been improved by the utilization of high-level synthesis (HLS) tools such as Vivado HLS. This tool allows to abstract from the detailed low-level knowledge of computer architecture, relying on automatic optimization mechanisms that interpret the high-level clauses inserted in the code. In our implementation, an extra layer of abstraction exists due to a directive-based programming model. OmpSs\cite{ompss} is developed at the Barcelona Supercomputing Center(BSC) and allows to engage the FPGAs just by adding pragma clauses in the code. At compilation time, OmpSs converts the code into an intermediate code interpreted by the Vivado HLS, simplifying the code writing to the scientific application experts.

CFD codes can profit the FPGAs for generating a customized architecture that can improve its performance per watt ratio. Early attempts\cite{NAG12, BUR15} of porting part of finite element method (FEM) codes have been tested successfully. Our work is based on a portable implementation method presented in ~\cite{OYA17}. The idea consists of implementing the CFD as a concatenation of sparse algebra operations. The primary function is the sparse matrix-vector multiplication (SpMV), and therefore it is the focus of attention in this paper. 

The SpMV is widely studied due to its importance in many scientific applications. Storage formats are used to avoid storing the zero elements of the sparse matrix. Consequently, the multiplying vector accesses are indirect and determined by the sparsity pattern of the matrix. The matrices arisen from the discretization over unstructured meshes have a very scarce sparsity pattern, making difficult the data reuse. Regarding the SpMV in FPGAs, different implementation strategies have been studied in the past.  An implementation based on using block structures present in the sparsity pattern of FEM matrices was proposed in \cite{GRI16}. A column-wise implementation was presented by ~\cite{DOR14}. The idea consists of avoiding the multiplying vector's indirect accesses, while the resulting vector can be stored in a dense local structure. Another strategy, developed in~\cite{GRI15}, is based on performing a lossless compression of the non-zero elements of the matrix for reducing the memory bandwidth occupancy. A cache vector strategy was presented in ~\cite{UMU15} and later utilized in\cite{UMU16}, the caching scheme aims at maximizing the data reuse of the multiplying vector by performing a preprocess that determines the cache misses that later work as an input for the algorithm. The cache misses have also been treated by fully transferring the multiplying vector into the BRAM~\cite{HOS19}. These works aim to be efficient in a broad set of matrices that does not profit the problem information of the sparsity pattern in our CFD matrices. 

Therefore, we propose an FPGA SpMV algorithm that borrows some of the developments mentioned above while incorporating information from the CFD matrices. A cache vector is implemented as a circular list for maximizing the memory reuse and concurrently compress part of the matrix data. The matrix is stored using a sliced ELLPACK format~\cite{{MON10}} following a column-wise order.

The rest of the paper is organized as follows. 
In the next section, we present the CFD simulations' application context to understand the importance of the SpMV.
In Section~\ref{sec:app} we present an overview of the hardware architecture and software components of the FPGA utilized in this paper. 
Section~\ref{sec:gen} describes the general considerations for implementating the SpMV.
Then, in Section~\ref{sec:spmv}, the proposed algorithm for the FPGA cache SpMV is explained in detail.
Afterward, in Section~\ref{sec:res}, the performance of the new algorithm is compared with the naive implementation and with a general-purpose implementation found in the literature.
Finally, we summarize our contributions in Section~\ref{sec:conclus}.

\section{Application Context}
\label{sec:app}

Our application context are the CFD simulations based on the solution of the Navier Stokes equations for incompressible turbulent flows:
\begin{eqnarray}
\diver {\vel} &=& 0, \label{ch3:NCont} \\
\frac{\partial {\vel}}{\partial t} + ( {\vel} \cdot \mnabla ) \vel  &=&
         -  \mnabla {p} +  \nu \lapl {\vel}  \label{ch3:NSeq}      
\end{eqnarray}
where ${\vel}$ is the three-dimensional velocity vector,
${p}$ is the kinematic pressure scalar field, and $\nu$ is the fluid's kinematic viscosity.

The standard strategy to discretize the equations is using stencils. Those operations aim at choosing a path to sweep the data maximizing its reuse. The data reuse is possible due to the geometrical topology in the structured grids. However, unstructured meshes lack the geometrical support to detect data re-utilization zones, and therefore, do not profit from the stencil implementation. In such cases, an alternative portable implementation based on algebraic operators has been presented in \cite{OYA13, ALV18}.

In an operator-based formulation, the finite-volume spatial discretization of these equations reads
\begin{eqnarray}
\mathbf{M} \vel_c &=\mathbf{0}_c, \label{ch3:OpCon} \\
\boldsymbol \Omega \frac{d\vel_c}{dt} + \mathbf{C}(\vels)\vel_c + \mathbf{D} \vel_c + \boldsymbol \Omega \mathbf{G} p_c &=\mathbf{0}_c, \label{ch3:OpMom}
\end{eqnarray}
where $\vel_c$ and $p_c$ are the cell-centered velocity and pressure fields,
$\mathbf{0}_c$ is the discrete collocated field with zero in each component,
$\vels$ is the velocity field projected to the faces' normals,
$\boldsymbol \Omega$ is a diagonal matrix with the sizes of control volumes,
$\mathbf{C}(\vels)$ and $\mathbf{D}$ are the convection and diffusion operators,
and finally, $\mathbf{M}$ and $\mathbf{G}$ are the divergence and gradient operators, respectively.

Note that $\boldsymbol \Omega$, $\mathbf{M}$ and $\mathbf{G}$ are linear operators, so they can be implemented as sparse matrices. The row $i$ of the matrix represents the discretization within a cell $i$. The non-zero columns within the row describe the connections with its neighboring cells. And the non-zero coefficients the relationship between the cells. On the other hand, the $\mathbf{C}(\vels)$ is a non-linear operator with coefficients that are updated each time-integration step using $\vels$, and therefore,  can not be mapped to a sparse matrix automatically.  The scalar field $\vels$  represents the velocity interpolated at the faces, living in $\mathbb{R}^{N_{s}}$
The solution consisted of creating an artificial operator that profits from the implementation knowledge\cite{OYA17} .
In practice, the coefficients of the convective term are stored in a one-dimensional array of dimension $N_{e}$, where
$N_{e}$ is the number of non-zero entries in $\mathbf{C}(\vels)$. The arrangement of this array depends on the storage format
chosen for the operator. Under these conditions, we define the evaluation of $\mathbf{C}(\vels)$ as a linear operator
$\mathbf{E_C}:\mathbb{R}^{N_s}~\mapsto \mathbb{R}^{N_e}$, such that:

\begin{align}%\label{ch3:eq:op_new}V
 \centering
 \mathbf{C}(\vels)   \equiv    \mathbf{E_C}  \vels.
\end{align}

\begin{figure}
    \centering
    \includegraphics[scale=.6]{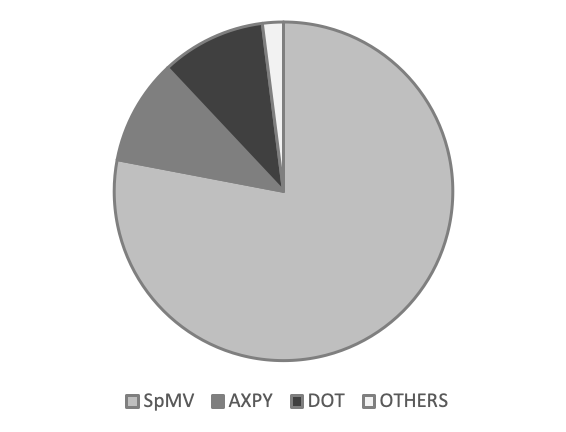}
    \caption{Main operations in CFD application developed under the portable implementation model.}
    \label{fig:queso}
\end{figure}

This strategy permits us to transform the non-linear operators into a concatenation of linear operators represented by sparse matrices. Consequently, the
time-integration step of the CFD simulation is a concatenation of few algebraic operations. In this work, we focus our attention on the SpMV that is the most time-consuming operation (78\%) from the linear solver and the matrix assembly when using the portable implementation~\cite{OYA17} according to Figure~\ref{fig:queso}. 

\begin{figure}[h]
    \centering
    \includegraphics[scale=.25]{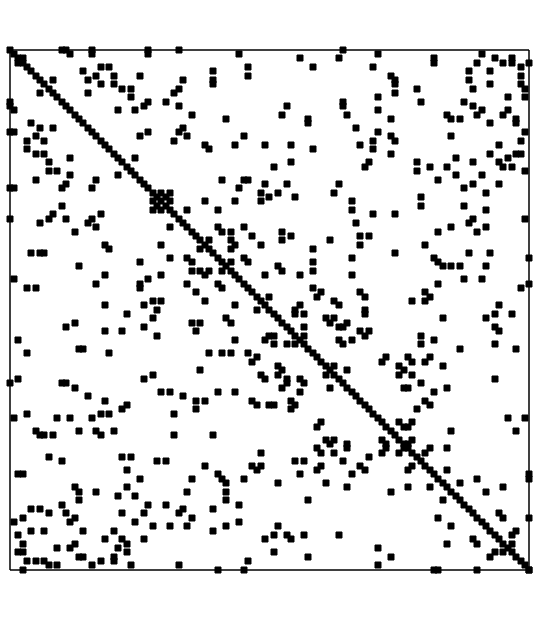}
    \caption{Sparsity pattern of the matrix for a tetrahedral mesh with 100 elements.}
    \label{fig:before}
\end{figure}

The SpMV is a memory bounded operation dominated by the indirect accesses of the multiplying vector. Those accesses are determined by the matrix's sparsity pattern, which in our simulations is derived from the discretization method and the connection of the unstructured mesh cells. For instance, a second-order discretization scheme derives in a matrix in which rows have up to five non-zero components on a mesh of tetrahedra. A CFD matrix typically has thousands of rows, leading to a very scarce sparsity pattern. This data path generally leads to low resource utilization and compute efficiency in modern architectures\cite{WIL09}. Figure ~\ref{fig:before} depicts the sparsity pattern of a matrix arisen from 100 tetrahedral cells using the mesh generator's numbering. Note that the matrix is symmetrical due to the discretization scheme.
In this work, we focus on matrices arisen from incompressible flow problems. In that case, the matrices remain constant during the simulation. This fact facilitates heavy pre-processing techniques since its relative cost would be negligible compared to the full simulation time.

\section{System overview}
\label{sec:cont}

\subsection{Computing resources}

The Field Programmable Gate Arrays can adapt the circuit logic to the algorithm that it is running on them. A configuration with an optimal number of computing gates and RAM blocks is possible, leading to an improvement in power consumption. 
In this work, we use the FPGA ADM-PCIE-7V3 of Xilinx that features two independent channels of DDR3 memory capable of 1333MT/s (fitted with two 8GB SODIMMs). Moreover, it consists of a BRAM of 52.9MB, FFs 886k, LUTs 693k, DSP3600. 

\begin{figure}[h]
    \centering
    \includegraphics[scale=.25]{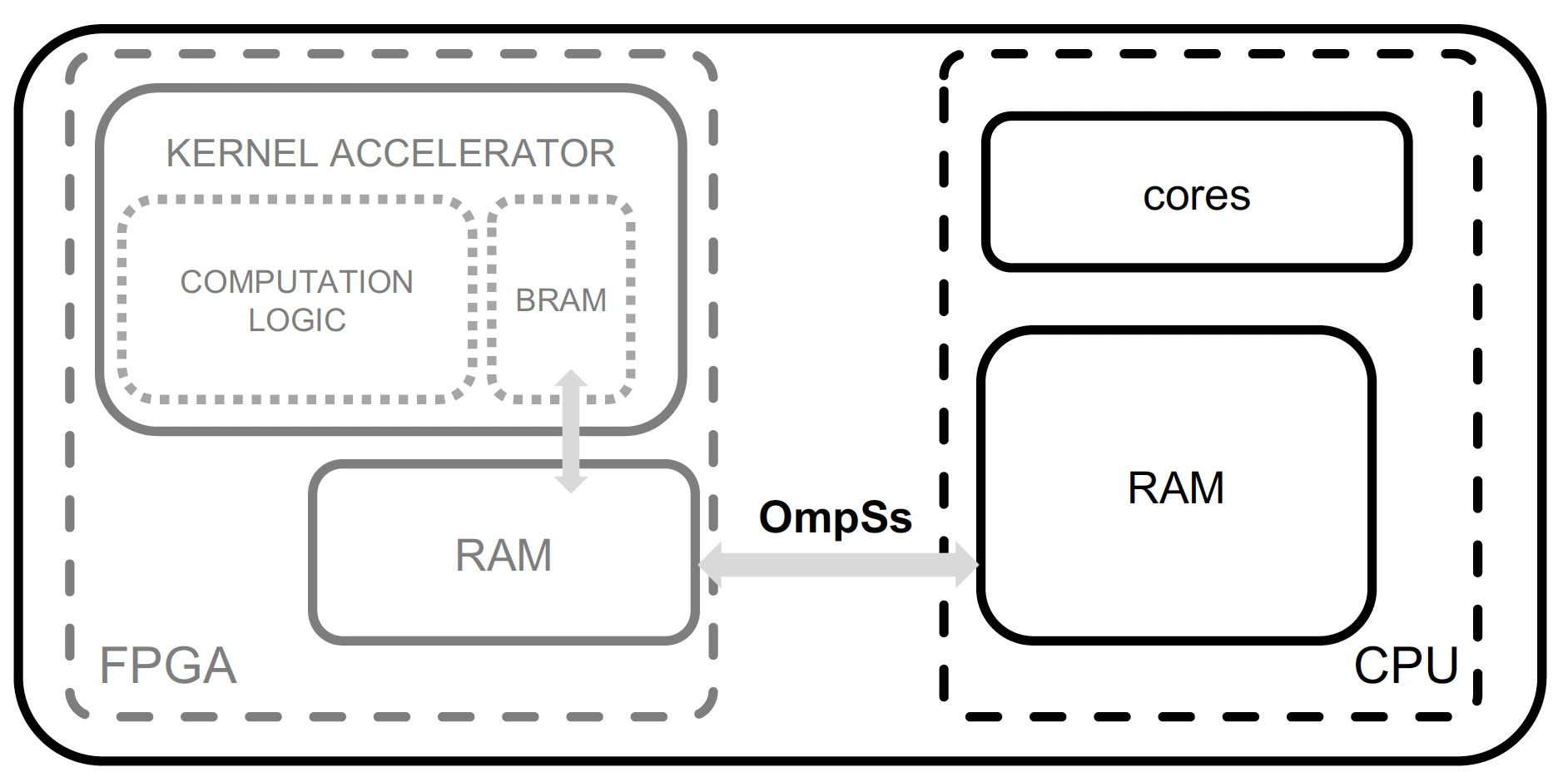}
    \caption{Node configuration diagram.}
    \label{fig:scheme}
\end{figure}

The FPGA is connected with a CPU through the PCI-express. Moving data into the FPGA can become a bottleneck since it needs to be transferred from the RAM of the node to the FPGA RAM. From there, it has to be transferred to the on-chip memory (BRAM) when processing the SpMV. A scheme with the node configuration is shown in Figure~\ref{fig:scheme}. This work assumes that the matrix data fit in the device's RAM and is transferred only at the simulation's preprocess. Therefore the memory transfers between CPU and FPGA RAMs are not considered.

\subsection{Software stack}

High-level synthesis (HLS) has facilitated the FPGA implementations during the last years.  It has abstracted the complex process of re-programming the FPGAs using the Register-Transfer Level (RTL) design and a Hardware Description Language (HDL).
Standard code is complemented with FPGA-specific directives that aim to explain the desired behavior and create digital hardware that maps such action.
Vivado HLS is the tool used by the Xilinx to perform the synthesis. In recent years, the HLS has been incorporated in directive-based programming models such as OmpSs. 
%By doing do, it combines the potential of using advanced tools for  taskification, or to just simplify the data transfer process.

In the OmpSs@FPGA programming model, programmers write a single source 
programs in C/C++ that run on heterogeneous systems with CPUs and FPGAs.
OmpSs follows an approach similar to OpenMP with some extensions to offload
tasks onto the FPGAs.

OmpSs programs are annotated with two types of directives:
\begin{itemize}
  \item \texttt{$\#$pragma omp task in(...) out(...) inout(...) }  are applied to function
prototypes, indicating that the invocations to such function are to be
spawned in parallel. Additionally, the input-output clauses (in, out, and inout) tell the data that needs to be available to the task.
And also, indicate if the data is consumed,
produced, or both. For example, this is the typical annotation for the sparse matrix
multiplication:

\begin{lstlisting}[language=C,basicstyle=\scriptsize]
#pragma omp task in([NNZ]A,[NNZ]cols,[N]x),inout([N]b) 
void spmv(double *A, int* cols, double* x, double *b);
\end{lstlisting}

 \item  \texttt{$\#$pragma omp target device(fpga)} is attached to a task to indicate that
the code of the function will be transferred to a separate output file to
be compiled with the Xilinx Vivado HLS tool, thus getting the bitstream
to configure the FPGA. Accordingly, the complete annotation  becomes:

\begin{lstlisting}[language=C,basicstyle=\scriptsize]
#pragma omp target device(fpga) 
#pragma omp task in([NNZ]A,[NNZ]cols,[N]x),inout([N]b)
void spmv(double *A, int* cols, double* x, double *b);
\end{lstlisting}

\end{itemize}

\begin{figure}[h]
    \centering
    \includegraphics[scale=.42]{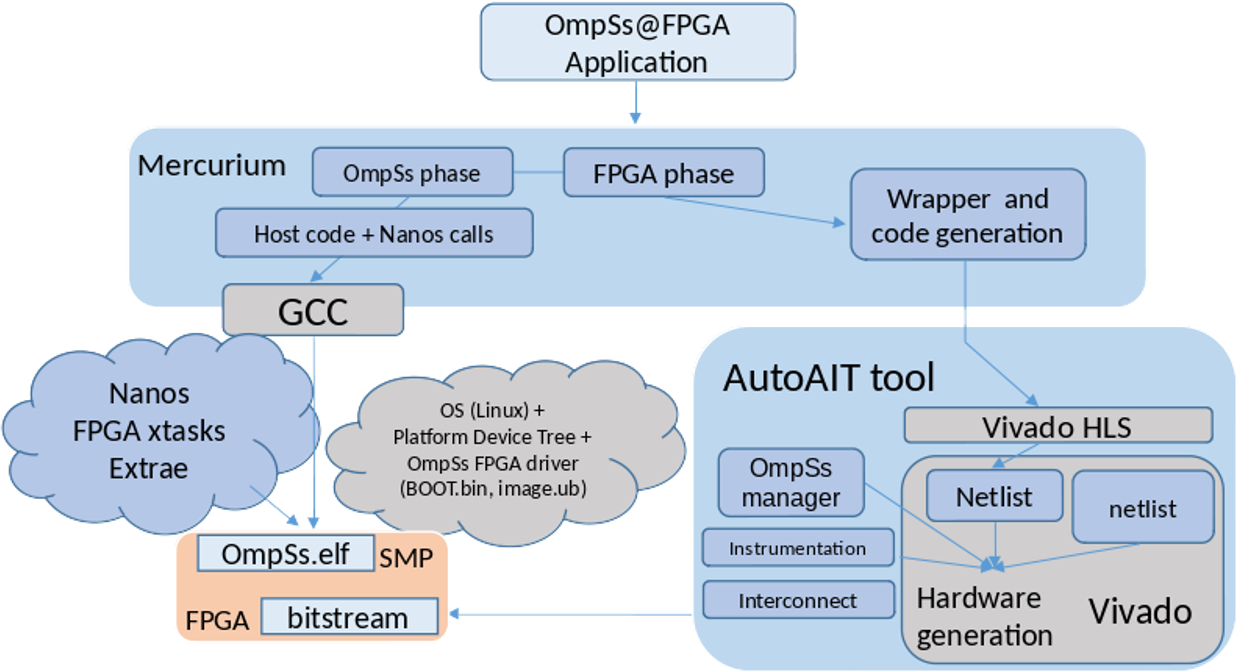}
    \caption{OmpSs@FPGA workflow.}
    \label{fig:ompss}
\end{figure}

This approach allows generating code for Xilinx FPGAs, from Zynq 7000, and
Zynq Ultrascale+, to the new PCI-Express attached ALveo boards.
The structure of the compilation environment is shown in Figure~\ref{fig:ompss}. Code
annotated with \texttt{target device(fpga)} is interpreted by Mercurium compiler to generate a Vivado HLS file.
This file is sent to the AutoAIT tool and compiled with Vivado HLS (right side of the figure). 
A bitstream is generated from the OmpSs IP
blocks with support to the task management tool. 
On the other hand, Standard C/C++ code for the CPUs is compiled with the native system compiler (GCC
in the left of the figure) and linked with the Nanos++ runtime system.  This runtime provides
the threading and tasking services managing the parallelism, data transfers, 
and the execution of the FPGA tasks.

In this work, we use OmpSs combined with Vivado HLS to design our algorithm. However, the algorithm itself does not depend on them and could be reused in any other language.

\section{General Considerations}
\label{sec:gen}

%This stage consists of three complementary operations that help reorganizing the data to make it suitable for the FPGA implementation.  

\subsection{Matrix Reordering}

\begin{figure}[h]
    \centering
    \includegraphics[scale=.25]{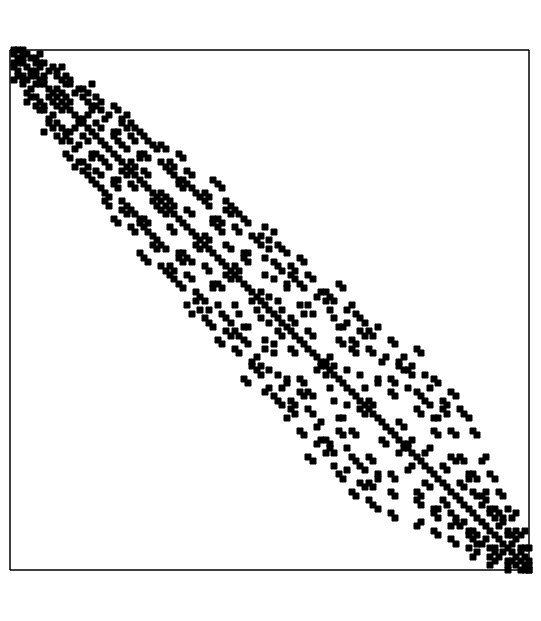}
    \caption{Sparsity pattern of the same matrix arisen from tetrahedral mesh (100 elements) after applying the Cuthill-Mckee reordering.}
    \label{fig:after}
\end{figure}

A Cuthill-Mckee algorithm~\cite{CUT69} is applied to reordering the unknowns of the system.  The sparsity pattern of the matrix is transformed into a band matrix. The maximum difference between column indexes of the same row is known as the width of the band. The reordering aims at reducing this width by applying a variant of a breath-first search of graph theory. The algorithm has proven to be efficient for CPU and GPU execution since it minimizes the cache misses produced by indirect memory accesses of the multiplying vector\cite{OYA17}. Figure~\ref{fig:after} shows the changes in the sparsity pattern of the same matrix with $100$ tetrahedra cells after applying the reordering. Note that the algorithm does not alter the matrix coefficients nor the sense of the equations. In practical terms, the matrix is still stored contiguously in memory, permitting it to be read efficiently. Moreover, the reordering is applied to the full CFD simulation, so its application is needed only once at the start of the simulation.

\subsection{Storage Format}
\begin{figure*}[ht!]
    \centering
    \includegraphics[scale=.33]{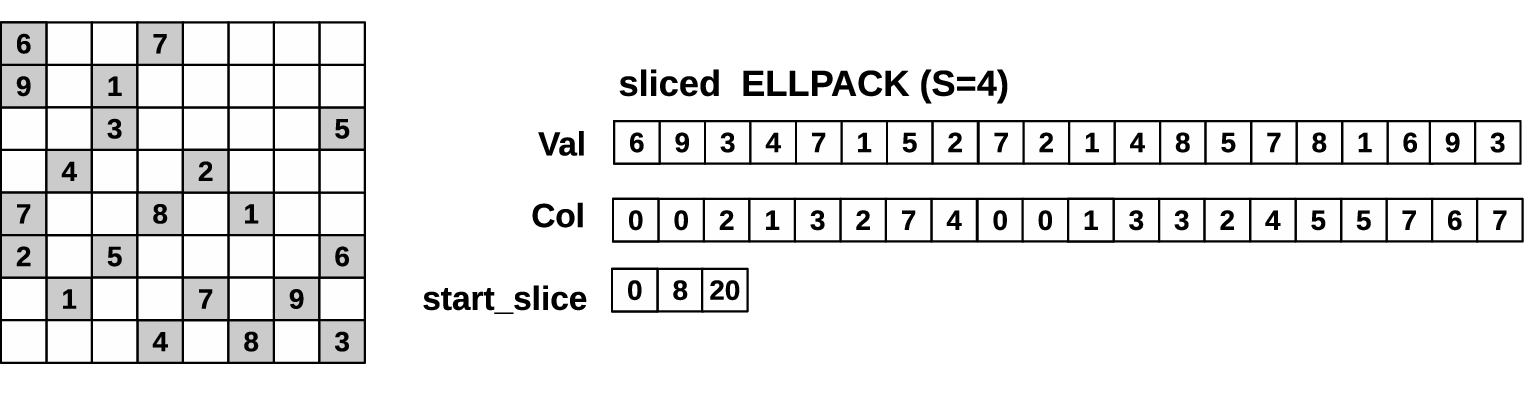}
    \caption{Example of the sliced ELLPACK format for a matrix.}
    \label{fig:sliced}
\end{figure*}
The sliced ELLPACK format~\cite{MON10} is the most appropriate for the matrices arisen from our CFD discretization~\cite{OYA17}. The format consists of dividing the matrix into slices with a similar number of non-zero elements per row. The number of rows per slice is a parameter defined as $S$. For each slice, the maximum number of non-zero components per row is calculated, and zeros are padded in the rows with fewer elements than the maximum. This scheme permits maintaining the regularity within each slice. Finally, each slice is stored using a column-wise order. By doing so, accessing the rows is performed independently and contiguously. This storage adapts better to our matrices in which the slices may represent different geometrical elements. Figure~\ref{fig:sliced} shows an example of the storage format when selecting a slice size four. In CFD, the array $Val$ contains double-precision matrix coefficients to conserve the accuracy of the simulation.  The array $col$ contains the column indexes in a 4-byte integer representation. 

\subsection{Naive implementation}

The primary difficulty in the SpMV is reading the multiplying vector. The matrix's column index, stored as an integer, needs to be read beforehand to determine the vector's required component. This dependency adds an overhead while reading the multiplying vector.
In a matrix derived from a tetrahedral mesh, each of the vector components is read on average five times. However, the reuse of the elements not necessarily takes place between two adjacent rows. Consequently, in a naive implementation, each vector component is read up to five times from global memory following an irregular pattern. 

%\textbf{Faltaría agregar el pseudo algorithm}

\section{FPGA SpMV for CFD using unstructured meshes}
\label{sec:spmv}
\subsection{Cache-like data structure}

The band matrix reordering provides a data path for accessing the multiplying vector that exposes the reusing of its components. As the algorithm moves through the matrix's rows, the multiplying vector elements are being reused until a point in which the elements are not needed anymore. 

Due to the banded pattern, the matrix is divided into slices of $S$ rows. $S$ is the optimal number of rows needed to hide the latency of pipelining them on the FPGA. 
For each slice $i$, the first and last vector components needed are obtained from the sparsity pattern and named $col\_start_i$ and $col\_end_i$ respectively. The two values determine the range of the vector components required by the slice $i$. When solving two adjacent slices, the necessary elements overlap, setting the basis for data reuse. Then, each IP core is implemented to operate on a block of $B$ adjacent slices executed in order.
In each block execution, the multiplying vector components needed by the first slice are loaded into a cache vector stored in BRAM (local memory). The next slices upload only the members of the multiplying vector that are not already in the cache. Load operations are performed in chunks of memory, so-called $wordsize$. For simplicity, the $wordsize$ is equal to the number of rows in a slice ($S$).

The cache vector is a circular list in which the new components are stored consecutively. When the end of the cache vector is reached, the new components are saved from the beginning, thus, erasing the older cached elements. The cache vector's optimal size is calculated from the width of the band obtained from the Cuthill-Mckee reordering.

The matrix's storage format needs two additional parameters: 1) the number of the words needed from the multiplying vector that are not currently in the cache vector, and 2) the position of the multiplying vector where the block needs to start the reading process. The number of components read by $block_i$ is obtained from the $colend_i$ – $colend_{i-1}$. And $offset_i$ is equal to $coldend_{i-1}$.

%\textbf{Creo que me dejo el como se mueve a través del vector de la cache}

Finally, the column indexes are rewritten as their relative position within the cache vector. By doing so, the indexes vary in the range between zero and the cache vector length. This size is much smaller than the multiplying vector's total size, and therefore the column indexes can be stored as short integers saving 2-bytes per component read. This compression reduces part of the overhead of reading the components. During the multiplication, the cache vector is still accessed indirectly. However, the memory now is within the BRAM, which is much faster. 

\subsection{Adapted ELLPACK FPGA cached-SpMV} 

The matrix is stored using an ELLPACK format with a constant number of non-zero entries per row. This number is defined by the geometry of the unstructured meshes and the discretization order. For a tetrahedral mesh, a constant number of five non-zero entries per row. Note that in the case with less connectivity, such as boundary conditions, zeros are padded to enforce the size.  This format has proven to be the most efficient in our matrices~\cite{OYA17}. In each BLOCK, the rows are stored using a column-wise order that permits the execution of all the rows' calculation concurrently.

Each instance of our SpMV executes a block of many slices, and each one performs three steps:
\begin{itemize}
    \item Load the words needed that are not in the cache vector. 
    \item Load matrix and solve the multiplication of each row of the slice.
    \item Copy the result back to the CPU.
\end{itemize}

The code is implemented using a combination of OmpSs and Vivado HLS.
\section{Numerical Results}
\label{sec:res}

The matrices used in the numerical experiments were generated using the Laplacian operator and a second-order discretization applied on cells of tetrahedral meshes. Table~\ref{tab1} shows the characteristics of those matrices. 

\begin{table}[htbp]
\caption{Characteristics of the matrices used in the numerical results}
\begin{center}
\begin{tabular}{|c|c|c|c|c|c|}
\hline
\textbf{} & \multicolumn{5}{|c|}{\textbf{Matrices}} \\
\cline{2-6} 
\textbf{Charac.} & \textbf{\textit{C50K}}& \textbf{\textit{C100K}}& \textbf{\textit{C200K}}& \textbf{\textit{C400K}}& \textbf{\textit{C800K}} \\
\hline
\textbf{rows} & $49,336$ & $97,521$ & $187,078$ & $398,000$ & $775,058$ \\
\textbf{nnz} & $243,504$ & $481,855$ & $ 928,184$ & $1,975,128$ & $3,851,840$ \\
\hline
\end{tabular}
\label{tab1}
\end{center}
\end{table}

\subsection{Comparison with a naive approach}
First, we have compared our algorithm with the naive approach discussed in Section~\ref{sec:gen}. Note that the naive SpMV IP has lower BRAM requirements since the multiplying vector is not cached. Therefore, up to 32 instances can be generated and executed concurrently on the FPGA. On the other hand, the optimal version of the cache SpMV has been found when using the number of rows per slice, $S$, equal to 512. The $wordsize$ is equal to the parameter $S$. A cache vector of 16,384 entries (32 words) is allocated in the BRAM. With these constraints, only four IPs of the cache SpMV can be instantiated and executed concurrently. The results for both executions are shown in Table~\ref{tab2}.

\begin{table}[htbp]
\caption{Comparison with the naive format}
\begin{center}
\begin{tabular}{|c|c|c|c|c|c|}
\hline
\textbf{} & \multicolumn{5}{|c|}{\textbf{Matrices}} \\
\cline{2-6} 
\textbf{Method} & \textbf{\textit{C50K}}& \textbf{\textit{C100K}}& \textbf{\textit{C200K}}& \textbf{\textit{C400K}}& \textbf{\textit{C800K}} \\
\hline
\textbf{Naive} & $0.138$ & $0.277$ & $0.545$ & $1.183$ & $2.335$ \\
\textbf{cached SpMV} & $0.017$ & $0.025$ & $ 0.036$ & $0.074$ & $0.142$ \\
\textbf{Speedup} & \textbf{$8.19$} & \textbf{$10.95$} & \textbf{$15.19$} & \textbf{$15.87$} & \textbf{$16.41$} \\
\hline
\end{tabular}
\label{tab2}
\end{center}
\end{table}

The comparison shows that the caching strategy outperforms the naive version up to 16 times. The acceleration grows with the matrix size since the multiplying vector's indirect accesses are more distanced.

\subsection{Comparison with an open library}

The next step consisted of comparing the results with an open library of the SpMV in FPGAs~\cite{HOS19}. The library implements a stream version of the SpMV in the traditional CSR format. The full multiplying vector is loaded once into the BRAM. By doing so, the indirect accesses are produced inside of a memory that is much faster. 
The drawback of following this approach is that much more BRAM is needed. Large matrix sizes are prohibited, or fill the BRAM with only one instance. Table~\ref{tab3} show the results for both executions. First, we have compared our algorithm with the naive approach discussed in Section~\ref{sec:gen}. Note that the naive SpMV IP has lower BRAM requirements since the multiplying vector is not cached. Therefore, up to 32 instances can be generated and executed concurrently on the FPGA.
On the other hand, the optimal version of the cache SpMV has been found when using the number of rows per slice, $S$, equal to 512. The $wordsize$ is equal to the parameter $S$. A cache vector of 16,384 entries (32 words) is allocated in the BRAM. With these constraints, only four IPs of the cache SpMV can be instantiated and executed concurrently. The results for both executions are shown in Table~\ref{tab2}.   

\begin{table}[htbp]
\caption{Comparison with an open library}
\begin{center}
\begin{tabular}{|c|c|c|c|c|c|}
\hline
\textbf{} & \multicolumn{5}{|c|}{\textbf{Matrices}} \\
\cline{2-6} 
\textbf{Method} & \textbf{\textit{C50K}}& \textbf{\textit{C100K}}& \textbf{\textit{C200K}}& \textbf{\textit{C400K}}& \textbf{\textit{C800K}} \\
\hline
\textbf{stream CSR}  & $0.112$ & $0.221$ & $0.424$ & $0.902$ & $1.757$ \\
\textbf{cached SpMV} & $0.017$ & $0.025$ & $ 0.036$ & $0.074$ & $0.142$ \\
\textbf{Speedup} & \textbf{$6.64$} & \textbf{$8.73$} & \textbf{$ 11.83$} & \textbf{$12.11$} & \textbf{$12.35$} \\
\hline
\end{tabular}
\label{tab3}
\end{center}
\end{table}

The comparison shows that the caching strategy outperforms the library version up to 12 times. The acceleration grows with the matrix size since the indirect accesses of the multiplying vector are more distanced. The $C800K$ was the last matrix size that the library could solve.

\section{Conclusions}
\label{sec:conclus}

The SpMV has proven to be a complex problem to be solved by FPGAs. Its difficulty relies on the indirect memory accesses of the multiplying vector. The sparsity pattern in the matrices arisen from unstructured CFD simulations is scarce, adding an extra layer of complexity. This work shows how applying some pre-processing algorithms tuned an optimal SpMV FPGA execution. An implementation of the SpMV using cache is proposed. The promising results show that using the cache version outperforms the naive implementation up to 16 times. Moreover, the cache SpMV has also been compared against a library that implements a strategy of caching the full multiplying vector into the BRAM. In this case, our implementations are up to 12 times faster, proving their potential to be used in future CFD simulations.

\section*{Acknowledgments}
The research leading to these results has received funding from the European Union Horizon 2020 Programme under the LEGaTO Project (www.legato-project.eu), grant agreement nº 780681. This work is also partially supported by the European Union's Horizon 2020 research and innovation programme under grant agreement number: 846139 (Exa-FireFlows).
This paper expresses the opinions of the authors and not necessarily those of the European Commission. 
The European Commission is not liable for any use that may be made of the information contained in this paper.

%\section*{References}


\begin{thebibliography}{00}

\bibitem{DON11} J. Dongarra el at., The international exascale software project roadmap, International Journal of High Performance Computing Applications 25 (1) (2011) 3–60, http://dx.doi.org/10.1177/1094342010391989

\bibitem{top500} Top500 list. http://top500.org 

\bibitem{OYA20} G. Oyarzun, I. A. Chalmoukis, G. A. Leftheriotis, A. A. Dimas, A GPU-based algorithm for efficient LES of high Reynolds number flows in heterogeneous CPU/GPU supercomputers, Applied Mathematical Modelling, Volume 85, 2020, Pages 141-156, https://doi.org/10.1016/j.apm.2020.04.010.

\bibitem{BOR20} R. Borrell, D. Dosimont, M. Garcia-Gasulla, G. Houzeaux, O. Lehmkuhl, V. Mehta, H. Owen, M. Vazquez, and G. Oyarzun, ´ “Heterogeneous CPU/GPU co-execution of CFD simulations on the POWER9 architecture: Application to airplane aerodynamics,” Future Generation Computer Systems, vol. 107, Jun. 2020.

\bibitem{ompss}OmpSs. https://pm.bsc.es/ompss-at-fpga

\bibitem{NAG12} Z. Nagy, C. Nemes, A. Hiba, A. Kiss, Á. Csík and P. Szolgay, "FPGA based acceleration of computational fluid flow simulation on unstructured mesh geometry," 22nd International Conference on Field Programmable Logic and Applications (FPL), Oslo, 2012, pp. 128-135, doi: 10.1109/FPL.2012.6339276.

\bibitem{BUR15} P. Burovskiy, P. Grigoras, S. Sherwin and W. Luk, "Efficient assembly for high order unstructured FEM meshes," 2015 25th International Conference on Field Programmable Logic and Applications (FPL), London, 2015, pp. 1-6, doi: 10.1109/FPL.2015.7293749.

\bibitem{OYA17}Oyarzun, Guillermo \& Borrell, R. \& Gorobets, Andrey \& Oliva, Assensi. (2017). Portable implementation model for CFD simulations. Application to hybrid CPU/GPU supercomputers. International Journal of Computational Fluid Dynamics. 31. 1-16. 10.1080/10618562.2017.1390084. 

\bibitem{GRI16} P. Grigoraş, P. Burovskiy, W. Luk and S. Sherwin, "Optimising Sparse Matrix Vector multiplication for large scale FEM problems on FPGA," 2016 26th International Conference on Field Programmable Logic and Applications (FPL), Lausanne, 2016, pp. 1-9, doi: 10.1109/FPL.2016.7577352.

\bibitem{DOR14} Richard Dorrance, Fengbo Ren, and Dejan Marković. 2014. A scalable sparse matrix-vector multiplication kernel for energy-efficient sparse-blas on FPGAs. In Proceedings of the 2014 ACM/SIGDA international symposium on Field-programmable gate arrays (FPGA ’14). Association for Computing Machinery, New York, NY, USA, 161–170. DOI:https://doi.org/10.1145/2554688.2554785

\bibitem{GRI15} Grigoras, Paul, Pavel Burovskiy, Eddie Hung and Wing-Pong Luk. “Improving SpMV Performance on FPGAs through Lossless Nonzero Compression.” (2015).

\bibitem{UMU15} Umuroglu Y., Jahre M. (2015) A Vector Caching Scheme for Streaming FPGA SpMV Accelerators. In: Sano K., Soudris D., Hübner M., Diniz P. (eds) Applied Reconfigurable Computing. ARC 2015. Lecture Notes in Computer Science, vol 9040. Springer, Cham

\bibitem{UMU16} Yaman Umuroglu, Magnus Jahre, Random access schemes for efficient FPGA SpMV acceleration, Microprocessors and Microsystems, Volume 47, Part B, 2016, Pages 321-332,
ISSN 0141-9331, https://doi.org/10.1016/j.micpro.2016.02.015

\bibitem{HOS19} Hosseinabady, M., \& Nunez-Yanez, J. (2019). A Streaming Dataflow Engine for Sparse Matrix-Vector Multiplication using High-Level Synthesis. IEEE Transactions on Computer-Aided Design of Integrated Circuits and Systems. https://doi.org/10.1109/TCAD.2019.2912923


\bibitem{MON10} A. Monakov, A. Lokhmotov, A. Avetisyan, Automatically Tuning Sparse Matrix-Vector Multiplication for GPU Architectures, High Performance Embedded Architectures and Compilers Lecture Notes in Computer Science Volume 5952, 2010, pp 111-125





\bibitem{OYA13} G. Oyarzun, R. Borrell, A. Gorobets, O. Lehmkuhl, A. Oliva, Direct Numerical Simulation of Incompressible Flows on Unstructured Meshes Using Hybrid CPU/GPU Supercomputers, Procedia Engineering, Volume 61, 2013, Pages 87-93, ISSN 1877-7058, https://doi.org/10.1016/j.proeng.2013.07.098.

\bibitem{ALV18}Alvarez Farre, Xavier \& Gorobets, Andrey \& Trias, F.Xavier \& Borrell, R. \& Oyarzun, Guillermo. (2018). $HPC^2$ —A fully-portable, algebra-based framework for heterogeneous computing. Application to CFD. Computers \& Fluids. 173. 10.1016/j.compfluid.2018.01.034. 

\bibitem{WIL09} S. Williams, L. Oliker, R. Vuduc, J. Shalf, K. Yelick and J. Demmel, "Optimization of sparse matrix-vector multiplication on emerging multicore platforms," SC '07: Proceedings of the 2007 ACM/IEEE Conference on Supercomputing, Reno, NV, USA, 2007, pp. 1-12, doi: 10.1145/1362622.1362674.


\bibitem{CUT69} E. Cuthill and J. McKee. Reducing the bandwidth of sparse symmetric matrices In Proc. 24th Nat. Conf. ACM, pages 157--172, 1969.






%\bibitem{b1} G. Eason, B. Noble, and I. N. Sneddon, ``On certain integrals of Lipschitz-Hankel type involving products of Bessel functions,'' Phil. Trans. Roy. Soc. London, vol. A247, pp. 529--551, April 1955.
%\bibitem{b2} J. Clerk Maxwell, A Treatise on Electricity and Magnetism, 3rd ed., vol. 2. Oxford: Clarendon, 1892, pp.68--73.
%\bibitem{b3} I. S. Jacobs and C. P. Bean, ``Fine particles, thin films and exchange anisotropy,'' in Magnetism, vol. III, G. T. Rado and H. Suhl, Eds. New York: Academic, 1963, pp. 271--350.
%\bibitem{b4} K. Elissa, ``Title of paper if known,'' unpublished.
%\bibitem{b5} R. Nicole, ``Title of paper with only first word capitalized,'' J. Name Stand. Abbrev., in press.
%\bibitem{b6} Y. Yorozu, M. Hirano, K. Oka, and Y. Tagawa, ``Electron spectroscopy studies on magneto-optical media and plastic substrate interface,'' IEEE Transl. J. Magn. Japan, vol. 2, pp. 740--741, August 1987 [Digests 9th Annual Conf. Magnetics Japan, p. 301, 1982].
%\bibitem{b7} M. Young, The Technical Writer's Handbook. Mill Valley, CA: University Science, 1989.
\end{thebibliography}
\end{document}